# MAGNET DESIGN ISSUES & CONCEPTS FOR THE NEW INJECTOR

P. Fabbricatore, INFN Sezione di Genova, Italy

*Abstract*

Possible layouts of superconducting dipoles for the main injector of High Energy LHC (HE-LHC) are proposed on the basis of the experience matured with ongoing R&D activities at the Italian National Institute of Nuclear Physics (INFN), targeted at developing the technologies for high field fast cycled superconducting magnets for the SIS300 synchrotron of FAIR. Two different magnets are analysed: a) a 4 T dipole ramped up to 1.5 T/s, and b) a 6 T dipole to be operated at lower ramp rates.

## INTRODUCTION

The Facility for Anti-proton and Ion Research (FAIR), under development at GSI, includes the synchrotron SIS300 [1]. The name of the accelerator is related to its 300 Tm magnetic rigidity, which is needed for bending high intensity proton beams (90 GeV) and heavy ions, e.g. $U^{92+}$ up to 34 GeV/u. The dipole magnets have to be pulsed from the injection magnetic field of 1.0 T up to 4.5 T maximum field, at the rate of 1 T/s. The lattice includes two kinds of dipoles, only differing in length (3.9 m and 7.8 m) [2]. These magnets have the same geometrical cross-section with $\cos(\theta)$ shaped coils, 100 mm bore and the particular characteristic to be geometrically curved, with a sagitta ranging from 28 mm for the short magnets to 112.9 mm for the long ones.

Since 2006, R&D activities are going on at the Italian National Institute of Nuclear Physics (INFN) aimed at developing the technologies for constructing these magnets. The activity is performed in the framework of a project called DISCORAP (*DIpoli SuperCOnduttori RApidamente Pulsati*), according to a specific INFN-FAIR Memorandum of Understanding signed by both institutions in December 2006.

Important steps of the DISCORAP project have been: a) the development of a low loss superconducting Rutherford cable [3], b) the construction of coil winding models for assessing the constructive feasibility of curved coils, c) the construction of a complete model magnet composed of a cold mass enclosed in its horizontal cryostat [4]. The last step is now close to be concluded.

The main parameters of the model magnet for SIS300 are shown in Table 1. The conductor involved in this magnet is similar to the cable used in the outer layer of the LHC main dipole. It is a 36-strand Rutherford cable optimized for low ac losses as discussed later. Some characteristics of strand and cable are reported in Table 2.

On the basis of this experience we try to give information and develop considerations aimed at addressing general and specific aspects of the dipole for the main injector of HE-LHC.

Hans Müller of GSI/FAIR is acknowledged for the fruitful discussions and for the revision of this paper

As starting point we assume that the protons are injected at 100 GeV and accelerated up to 1 TeV or, at maximum, to 1.5 TeV, hence involving a 4 T dipole ramped up from 0.4 T, and a 6 T dipole, respectively. For the field rates we considered values in the range of 1÷1.5 T/s.

There are two critical aspects concerning these dipoles. The first one is of mechanical nature, since the magnets have to support $10^7$ magnetic cycles [5]. The second one is related to the need to limit the coil heating and reduce efficiently the heat dissipation [6]. The mechanical issues and the heat exchange problematic are related to the winding (lay-out, manufacture), the aspects of the heat dissipation are more related to the conductor design.

Table 1: Characteristics of the SIS300 model dipole under development at INFN

| Parameter | Value |
|---|---|
| Magnetic Field (T) | 4.5 |
| Ramp rate (T/s) | 1 |
| Coil aperture (mm) | 100 |
| Magnetic length (mm) | 3879 |
| Maximum operating temperature (K) | 4.7 |
| Layers/Turns per quadrant | 1/34 in 5 blocks (17,9,4,2,2) |
| Operating current (A) | 8920 |

Table 2: Characteristics of the cable used in the SIS300 model dipole

| | |
|---|---|
| Strand diameter (mm) | 0.825 |
| Filament twist pitch (mm) | 5 |
| Strand Ic @ 5 T, 4.22 K | >541 |
| n-index @ 5 T, 4.22 K | >30 |
| Stabilization matrix | Pure Cu and CuMn |
| Strand Number | 36 |
| Cable width (mm) | 15 |
| Cable thickness, thin edge (mm) | 1.362 |
| Cable thickness, thick edge (mm) | 1.598 |
| Transposition pitch (mm) | 100 |

## TEMPERATURE MARGIN

For any superconducting magnet the temperature margin is an important parameter. For a magnet operating in ac mode, it is a key parameter because the heat load due to the ac losses causes an increase of the coil temperature, predictable only with some uncertainties and depending on parameters difficult to be fully controlled. For the SIS300 dipole we designed a temperature margin of 1 K, which is presently reduced to 0.75 K because the developed low loss conductor has a critical current 14% lower than specified. Furthermore we computed that the ac losses cause a (local) temperature increase of up to 0.25 K. The real margin is consequently reduced to 0.5 K.

The temperature margin is given by the difference between the current sharing temperature and the operating temperature. Let be $I_c(B,T)$ the function describing how the critical current of the conductor depends on the magnetic field and temperature [7], and $I(B)=\alpha B$ the magnet load line identifying the peak field in the winding. The current sharing temperature $T_g$ is univocally indentified by the intersection of $I_c(B, T)$ with the load line at the operating current. The problem with this definition is that the functions involved can not be inverted for giving an analytical expression of $T_g$. Therefore we will use for the margin the definition given by M. Wilson [8]:

$$\Delta T = T_g - T_0 = [T_c(B) - T_0]\left[1 - \frac{I_0}{I_c(B,T_0)}\right], \quad (1)$$

which is valid for a linear dependence of the critical current on the temperature. In Eq. 1 $I_0$ is the operating current, $T_0$ the operating temperature and $T_c(B)$ the critical temperature as function of the magnetic field:

$$T_c(B) = T_{c0}\left[1 - \frac{B}{B_{c20}}\right]^{1/1.7}, \quad (2)$$

where $T_{c0}$ is the critical temperature (9.2 K for NbTi) and $B_{c20}$ is the critical field (14.5 T for NbTi).

From Eqs. 1 and 2 we can find a very simple expression relating the ratio of operating current critical current at fixed field and the temperature margin $\Delta T$:

$$f = \frac{I_0}{I_c(B,T_0)} = 1 - \frac{\Delta T}{T_{c0}\left[1 - \frac{B}{B_{c20}}\right]^{1/1.7} - T_0}. \quad (3)$$

In Fig.1 this function is plotted vs. the magnetic field for two different values of the temperature margin (0.5 K and 1 K), allowing to make some interesting considerations about the margin in current we have to take. As nominal temperature we have assumed $T_0$=4.7 K coming from SIS300 parameters. The magnetic field in the abscissa is the peak field. For a dipole generating 4 T field (peak field of about 4.4÷4.5 T) we have to work at 64% of the critical current at fixed field for a margin of 1 K and at 82% for 0.5 K margin. A 6 T magnet (peak field presumably about 6.4 T) requires to be operated at 45% of the critical current for 1 K margin and 72% for 0.5 K margin. The critical issue here is the amount of superconducting material required. For a 6 T magnet operating with 1 K margin we have to check if a real winding can be fitted in.

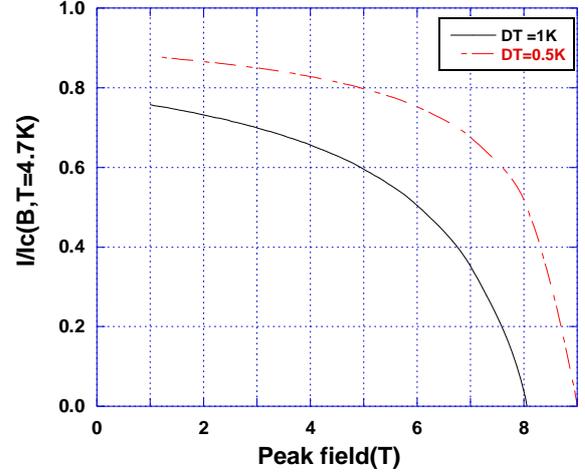

Figure 1: Operating to critical current ratio as function of the peak magnetic field for two different values of the temperature margin.

With this aim, let us try to evaluate the number of layers involved in a 4 T and 6 T dipole. For sake of simplicity we consider a sector coil [9] (just made of one sector) producing a dipole field $B$, which is directly proportional to the overall current density $J_{ov}$ and the radial thickness of the sector $w$:

$$B = \frac{\mu_0}{\pi}\sqrt{3}J_{ov}w, \quad (3)$$

considering that $B_{peak} = \gamma B$, we can find an expression for the sector thickness

$$w = \frac{\pi}{\mu_0\sqrt{3}}\frac{B_{peak}}{\gamma f J_c(B_{peak},T_0)\xi}. \quad (4)$$

For our calculations we use $\xi$ (the fraction of superconductor in the winding)=0.283, $\gamma$ (the ratio peak magnetic field to central field)=1.09 and $J_c(B, T = 4.7K)$ as calculated with a Bottura fit [7]. The results are shown in Fig.2. A dipole magnet producing a field of 4 T requires a coil radial thickness of 13÷14 mm for a temperature margin of 1 K. For the same margin a 6 T coil must have a thickness of more than 50 mm or 30 mm for 0.5 K margin. In term of layers made of practical Rutherford cables, a 4 T dipole magnet involves only one layer, whilst a 6 T dipole requires 2 layers and the temperature margin is closer to 0.5 K than 1 K.

## PROPOSED MAGNETS

On the basis of the conclusions of the previous sections, the proposed option for 1 TeV maximum energy is a 4 T dipole composed of one layer. This magnet would be very similar to the SIS300 model under development at INFN. It is proposed to hold this lay-out except for the geometrical curvature. Consequently the characteristics for this option are the ones reported in Table 1 with the exclusion of the ramp rate (here 1.5 T/s) and the magnetic length.

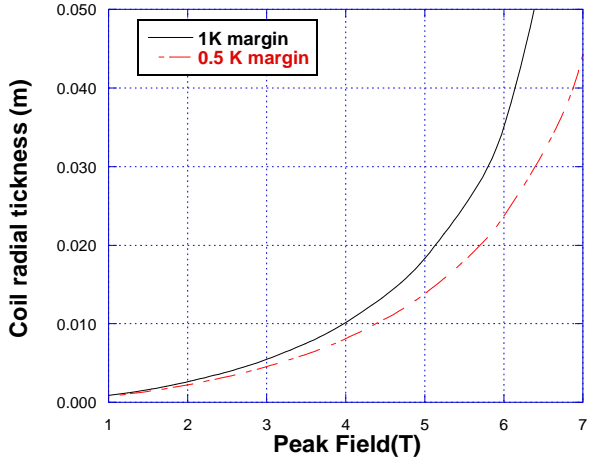

Figure 2: Coil radial thickness as function of the peak magnetic field for two different values of the temperature margin.

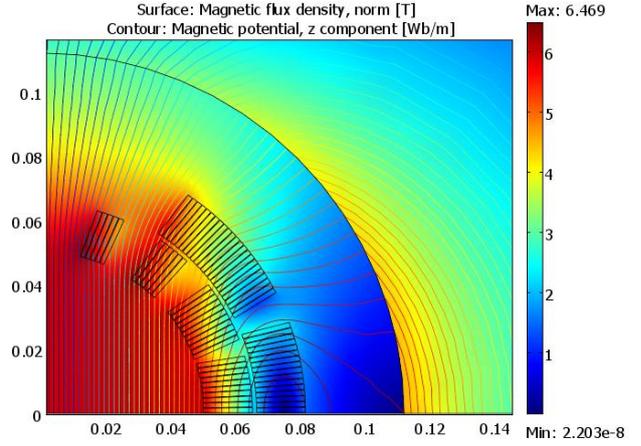

Figure 3: Layout of 6 T magnet (based on IHEP design) with the magnetic field distribution. The first quadrant is shown. The peak field is 6.42 T. The axes report dimensions in m.

For 1.5 TeV maximum energy we need a two layer coil. A very good candidate is the 6 T dipole developed at IHEP for SIS300 [10]. This design has been revisited and is proposed here with the characteristics shown in Table 3. The conductor is the same as for the 4 T option.

Table 3: Characteristics of the proposed 6 T option based on the 6 T SIS300 model dipole developed at IHEP

| Parameter | Value |
| --- | --- |
| Magnetic Field (T) | 6 |
| Ramp rate (T/s) | 1 |
| Coil aperture (mm) | 100 |
| Maximum operating temperature (K) | 4.7 |
| Layers/Turns per quadrant | 2/16 for first layer – 19 second |
| Operating current (A) | 6720 |

In Fig. 3 a cross section of the first quadrant of the magnet is shown, with the magnetic field distribution at the operating current. Only the winding and the iron are included.

## AC LOSSES

There are many sources of ac losses to be considered. They can be divided into three main categories: 1) ac losses in the conductor; 2) losses due to eddy currents in the mechanical structures; 3) losses in the iron yoke (magnetic, eddy and anomalous). Regarding the conductor, two main mechanisms are present: the hysteretic losses due to persistent currents in the filaments and the losses due to the intra-strand and inter-stand coupling currents.

The conductor and the magnet design of SIS300 were optimised for very low losses. The ac losses due to the persistent currents in the superconducting filaments were minimised using very fine filaments (2.5 μm geometrical diameter, 3.0 μm effective). The intra-strand coupling currents were minimised through both a small twist pitch (5 mm) and an optimised transverse electrical resistivity (0.44 nΩ). The inter-strand coupling currents were controlled through the contact resistance $R_a$ between adjacent strands. Our design value of $R_a$ is 200 μΩ. The contact resistance between opposite strands $R_c$ is very high (mΩ range) because a 25 μm thick stainless steel sheet has been inserted inside the Rutherford cable; i.e. we are using a cored cable [11].

Presently four lengths of low loss conductor have been produced at Luvata Pori (FI) under INFN contract. The characteristics of this cable are acceptable but not completely fulfilling requirements. The filament effective diameter is 3.0 μm as expected but the measured inter-strand resistivity is lower (0.3 nΩ) and the inter-strand resistance $R_a$ is higher than expected [12]. The average critical current of the extracted strand is 442 A (5 T, 4.22 K), or -14% compared to the design value. The critical current shows a large degradation of 6% after cabling and the n-index is 20. However, as stressed in [12], a new wire, with an improved design and an optimized manufacture cycle, is now under development at Luvata Pori.

The losses in the mechanical structure were reduced through the use of laminated collars: 3 mm thick austenitic plates electrically insulated. Steel laminations with a low value of the coercitive field ($H_c$= 40 A/m) were used for the yoke. The steel plates (1 mm thick) were electrically insulated and assembled using insulated bars.

Table 4 shows the different contributions to ac losses for the model of SIS300 magnet. The losses are given both in W/m and as percentage of the total power dissipation in ramping condition. The energy dissipated during a cycle will depend on the peculiarities of the cycle (time for ramp-up, flat top, ramp down and flat at injection field). The values in Table 4 are design values. After cable production, we are expecting a reduction of hysteresis losses in filaments, an increase of intra-strand coupling losses and a decrease of inter-strand losses with respect to these values.

Table 4: Calculated ac losses in the magnet body (losses in coil ends not included) for the INFN model of short dipole for SIS300 when ramping from 1.5 to 4.5 at 1 T/s.

| Loss source | Loss (W/m) | Loss fraction (%) |
|---|---|---|
| Hysteresis in sc filaments | 2.31 | 30 |
| Strand coupling | 0.69 | 9 |
| Interstrand coupling $R_a R_c$ | 0.46 | 6 |
| Eddy currents in collars, yoke and coil protection sheets | 0.46 | 6 |
| Yoke magnetic | 1.85 | 24 |
| Beam pipe | 1.08 | 14 |
| Collar connection elements (keys, pins) | 0.62 | 8 |
| Yoke connection elements (clamps, bars) | 0.23 | 3 |
| Total | 7.7 | 100 |

The same exercise done for the 4 T option is shown in Table 5. Computations were done for both 1 T/s and 1.5 T/s ramp rates. The information regarding the 6 T option is shown in Table 6.

Tables 5 and 6 report the ac losses for the two options on the basis of the present technology. In fact there are margins for further improvement requiring specific R&D activities. First of all it is necessary to improve the filament quality. The goal is an higher critical current density $J_c(5\ T, 4.22\ K)$=3000 A/mm$^2$, with filaments of effective diameter 2 μm. It is also important to better control the transverse resistivity through a manufacturing process limiting the filament deformation [12]. The strand twist pitch can be further reduced. The measurements done during the development demonstrated that a wire with diameter 0.825 mm could be twisted with a pitch as low as 4 mm, without a significant degradation of the critical current.

The use of electrical steel with lower coercitive field (30 A/m) can further decrease the contribution of the steel magnetization to the ac losses. Coil protection sheets made of insulating material can cut eddy currents in these components. There are also margins for decreasing the eddy currents in the other mechanical components. Table 7 reports the expected ac losses for the two proposed magnets after improving the conductor, the components and the design.

Table 5: Calculated ac losses in the magnet body (losses in coil ends not included) for the 4 T option when ramping from 0.4 to 4.0 T at different ramp rates.

| Loss source | Loss (W/m) and fraction 1 T/s | Loss (W/m) and fraction 1.5 T/s |
|---|---|---|
| Hysteresis in sc filaments | 3.11 (38%) | 4.65 (30%) |
| Strand coupling | 0.74 (9%) | 1.70 (11%) |
| Interstrand coupling $R_a R_c$ | 0.50 (6%) | 1.09 (7%) |
| Eddy currents in collars, yoke and coil protection sheets | 0.50 (6%) | 1.09 (7%) |
| Yoke magnetic | 1.57 (19%) | 2.63 (17%) |
| Beam pipe | 0.92 (11%) | 2.17 (14%) |
| Collar connection elements (keys, pins) | 0.67 (8%) | 1.55 (10%) |
| Yoke connection elements (clamps, bars) | 0.25 (3%) | 0.62 (4%) |
| Total | 8.26 | 15.50 |

Table 6: Calculated ac losses in the magnet body (losses in coil ends not included) for the 6T option when ramping from 0.4 to 6.0 T at 1 T/s

| Loss source | Loss (W/m) | Loss fraction (%) |
|---|---|---|
| Hysteresis in sc filaments | 5.40 | 40 |
| Strand coupling | 1.22 | 9 |
| Interstrand coupling $R_a R_c$ | 1.22 | 9 |
| Eddy currents in collars, yoke and coil protection sheets | 0.54 | 4 |
| Yoke magnetic | 3.10 | 23 |
| Beam pipe | 1.07 | 8 |
| Collar connection elements (keys, pins) | 0.68 | 5 |
| Yoke connection elements (clamps, bars) | 0.27 | 2 |
| Total | 13.5 | 100 |

The conductor ac losses in Tables 4÷7 were computed using Roxie™. The losses in the electrical steel were computed with FEMM [13]. Other computations were done with Comsol™. It is worth noting that the two options have very similar overall ac losses (about 11 W/m) and also the contributions to the losses are very similar. In all case there is a large contribution of persistent currents in the superconducting filaments (from 34% to 40%) and steel magnetization (from 20% to 25%)

Table 7: Calculated ac losses for 4 T dipole ramped at 1.5 T/s and 6 T ramped at 1 T/s (losses in coil ends not included).

| Loss source | Loss (W/m) and fraction 4 T dipole 1.5 T/s | Loss (W/m) and fraction 6 T dipole 1 T/s |
|---|---|---|
| Hysteresis in sc filaments | 3.91 (34%) | 4.24 (40%) |
| Strand coupling | 0.81 (7%) | 1.17 (11%) |
| Interstrand coupling $R_a$ $R_c$ | 0.92 (8%) | 0.85 (8%) |
| Eddy currents in collars, yoke and coil protection sheets | 0.11 (1%) | 0.11 (1%) |
| Yoke magnetic | 2.30 (20%) | 2.65 (25%) |
| Beam pipe | 2.18 (19%) | 1.06 (10%) |
| Collar connection elements (keys, pins) | 0.92 (8%) | 0.32 (3%) |
| Yoke connection elements (clamps, bars) | 0.35 (3%) | 0.21 (2%) |
| Total | 11.50 | 10.61 |

## A COMPARISON BETWEEN THE TWO OPTIONS

Table 8 shows a comparison of characteristics and performances for the two proposed options. The parameters considered for the comparison are: 1) the injection field and the sextupole component of the field, 2) the maximum and the peak magnetic fields, 3) the temperature margin over the maximum operating temperature of 4.7 K; 4) the AC losses in the superconducting cable during ramp; 5) the AC losses in the structures during ramp: eddy currents and magnetization; 6) the weight; 7) the construction costs.

Table 8: Comparison between 4 T and 6 T options for He-LHC main injector

| Parameter | 4 T dipole 1.5 T/s | 6 T dipole 1 T/s |
|---|---|---|
| Injection magnetic field [T] and b3 | 0.4 /-4.5 | 0.4 /-4.9 |
| Maximum/ Peak magnetic field [T] | 4/4.4 | 6/6.42 |
| Temperature Margin (K) over 4.7K | 1.66 | 0.65 |
| AC losses in the superconducting cable during ramp [W/m] | 5.6 | 6.3 |
| AC losses in the structures during ramp(eddy currents and magnetization) [W/m] | 5.9 | 4.3 |
| Weight (t/m) | 1.28 | 1.68 |
| Construction costs in (k€/m) | 60÷70 | 80÷90 |

Critical points for both magnets are the high values of the sextupole at the injection field. The 6 T option also works with a low temperature margin. Next year, the 6 T short dipole developed at IHEP, should be completely tested at GSI and the real limits would be clearer. The same considerations apply for the 4.5T model developed at INFN.

## CONCLUSIONS

The R&D developments for SIS300 dipoles both at INFN and at IHEP in collaboration with GSI are setting the basis for giving the feasibility of superconducting magnets with fields of 4.5÷6 T ramped at 1 T/s or faster.

Advanced designs, construction techniques and first low loss conductors were developed.

For more conclusive considerations we have to wait for results of the testing of the model magnets at operating temperatures at GSI next year. In particular we are waiting for more information regarding the effects due to mechanical fatigue, which could be a major problem for fast cycled magnet.

On the basis of the present knowledge some extrapolations can be done for HE LHC injector magnets. A 4 T dipole ramped at 1.5 T/s has been analysed and compared with a 6 T dipole to be operated at 1 T/s ramp rate.

It appears that one can get ac losses as low as 11 W/m when ramping the magnets. For a further reduction of the ac losses major variations of the design are required. The 4 T option is less critical and less expensive as the 6 T one.

The field quality at injection energy could be an issue for both options.

## REFERENCES

[1] W. F. Henning, "The GSI project: An international facility for ions and antiprotons",.Nuclear physics A, Nuclear ad hadronic physics 734 (2004) 654.
[2] G.Moritz, "Superconducting magnets, for the International accelerator facility for beams of ions and antiprotons at GSI" IEEE Trans Appl Supercond 13 (2003) 1329.
[3] G. Volpini, F. Alessandria, G. Bellomo, P. Fabbricatore, S. Farinon, U. Gambardella, M. Sorbi, "Low-loss NbTi Rutherford Cable for Application to the Development of SIS-300 Dipoles," IEEE Trans Appl Supercond 18 (2008) 997.
[4] P.Fabbricatore, F. Alessandria, G. Bellomo, S. Farinon, U. Gambardella, J.Kaugerts, R.Marabotto, G.Möritz, M. Sorbi, and G. Volpini, "Development of a curved fast ramped dipole for FAIR SIS300" IEEE Trans Appl Supercond 18 (2008) 232.
[5] S.Farinon, P.Fabbricatore, R.Musenich, F.Alessandria, G.Bellomo, M.Sorbi, G.Volpini, U.Gambardella, "A Model Dipole for FAIR SIS300: Design of the Mechanical Structure", IEEE Trans Appl Supercond 19 (2009) 1141.

[6] M. Sorbi, F. Alessandria, G. Bellomo, P. Fabbricatore, S. Farinon, U. Gambardella and G. Volpini, "Field Quality and Losses for the 4.5 T Superconducting Pulsed Dipole of SIS300" IEEE Trans Appl Supercond 18 (2008) 138.
[7] L.Bottura, "A practical fit for the critical surface of NbTi", IEEE Trans Appl Supercond 10 (2000) 1054.
[8] M.N.Wilson, "Superconducting Magnets", Oxford University Press, New edition (1987).
[9] L.Rossi and E.Todesco "Electromagnetic design of superconducting dipoles based on sector coils", Phys. Rev. ST Accel. Beams 10 (2007) 112401.
[10] S.Kozub, I.Bogdanov, V.Pokrovsky, A.Seletsky, P.Shcherbakov, L.Shirshov, V.Smirnov, V.Sytnik, L.Tkachenko, V. Zubko, E. Floch, G. Moritz, H.Mueller," SIS 300 Dipole Model", IEEE Trans Appl Supercond 20 (2010) 200.
[11] J. Kaugerts, G. Moritz, M. N. Wilson, A. Ghosh, A. den Ouden, I. Bogdanov, S. Kozub, P. Shcherbakov, L. Shirshov, L. Tkachenko, D. Richter, A. Verweij, G. Willering, P. Fabbricatore, and G. Volpini, "Cable Design for FAIR SIS 300", IEEE Trans Appl Supercond 17 (2007) 1477.
[12] G. Volpini, F. Alessandria, G. Bellomo, P. Fabbricatore, S. Farinon, U. Gambardella, M. Holm, B. Karlemo, R. Musenich, M. Sorbi, "Low loss NbTi superconducting Rutherford cable manufacture for the SIS300 INFN model dipole", contributed paper to ASC 2010, August 1-6, Washington D.C. (2010).
[13] Information about this code for "Finite Element Method Magnetics" can be foud at web site http://www.femm.info/wiki/HomePage